\input harvmac
\input amssym

\def\det{{\rm det}}


\def\IL{\relax{\rm I\kern-.18em L}}
\def\IH{\relax{\rm I\kern-.18em H}}
\def\IR{\relax{\rm I\kern-.18em R}}
\def\IC{\relax\hbox{$\inbar\kern-.3em{\rm C}$}}
\def\IZ{\relax\ifmmode\mathchoice
{\hbox{\cmss Z\kern-.4em Z}}{\hbox{\cmss Z\kern-.4em Z}}
{\lower.9pt\hbox{\cmsss Z\kern-.4em Z}} {\lower1.2pt\hbox{\cmsss
Z\kern-.4em Z}}\else{\cmss Z\kern-.4em Z}\fi}


\def\det{{\rm det}}

\font\manual=manfnt \def\dbend{\lower3.5pt\hbox{\manual\char127}}

\def\IZ{\relax\ifmmode\mathchoice
{\hbox{\cmss Z\kern-.4em Z}}{\hbox{\cmss Z\kern-.4em Z}}
{\lower.9pt\hbox{\cmsss Z\kern-.4em Z}} {\lower1.2pt\hbox{\cmsss
Z\kern-.4em Z}}\else{\cmss Z\kern-.4em Z}\fi}
\def\half {{1\over 2}}

\def\bar{\overline}

\def\rt2{\sqrt{2}}
\def\irt2{{1\over\sqrt{2}}}

\lref\bertolini{
  M.~Bertolini and M.~Trigiante,
  ``Microscopic entropy of the most general four-dimensional BPS black  hole,''
  JHEP {\bf 0010}, 002 (2000)
  [arXiv:hep-th/0008201].
}

\lref\cardoso{G.~Lopes Cardoso, B.~de Wit and T.~Mohaupt,
  ``Corrections to macroscopic supersymmetric black-hole entropy,''
  Phys.\ Lett.\ B {\bf 451}, 309 (1999)
  [arXiv:hep-th/9812082].}

\lref\osv{H.~Ooguri, A.~Strominger and C.~Vafa,
  ``Black hole attractors and the topological string,''
  Phys.\ Rev.\ D {\bf 70}, 106007 (2004)
  [arXiv:hep-th/0405146].}

\lref\dab{A.~Dabholkar,
  ``Exact counting of black hole microstates,''
  [arXiv:hep-th/0409148].}

\lref\erik{E.~Verlinde,
  ``Attractors and the holomorphic anomaly,''
  arXiv:hep-th/0412139;
}

\lref\DVV{
  R.~Dijkgraaf, E.~Verlinde and H.~Verlinde,
  ``Counting dyons in N = 4 string theory,''
  Nucl.\ Phys.\ B {\bf 484}, 543 (1997)
  [arXiv:hep-th/9607026].
}

\lref\finn{
  V.~Balasubramanian and F.~Larsen,
  ``On D-Branes and Black Holes in Four Dimensions,''
  Nucl.\ Phys.\ B {\bf 478}, 199 (1996)
  [arXiv:hep-th/9604189].
}

\lref\vijay{
  V.~Balasubramanian,
  ``How to count the states of extremal black holes in N = 8 supergravity,''
  arXiv:hep-th/9712215.
}

\lref\msw{
  J.~M.~Maldacena, A.~Strominger and E.~Witten,
  ``Black hole entropy in M-theory,''
  JHEP {\bf 9712}, 002 (1997)
  [arXiv:hep-th/9711053].
}

\lref\GaiottoGF{
  D.~Gaiotto, A.~Strominger and X.~Yin,
  ``New connections between 4D and 5D black holes,''
  arXiv:hep-th/0503217.
}

\lref\ssy{
  D.~Shih, A.~Strominger and X.~Yin,
  ``Recounting dyons in N = 4 string theory,''
  arXiv:hep-th/0505094.
}

\lref\kallk{
  R.~Kallosh and B.~Kol,
  ``E(7) Symmetric Area of the Black Hole Horizon,''
  Phys.\ Rev.\ D {\bf 53}, 5344 (1996)
  [arXiv:hep-th/9602014].
}

\lref\CremmerUP{
  E.~Cremmer and B.~Julia,
  ``The SO(8) Supergravity,''
  Nucl.\ Phys.\ B {\bf 159}, 141 (1979).
}

\lref\GaiottoXF{
  D.~Gaiotto, A.~Strominger and X.~Yin,
  ``New connections between 4D and 5D black holes,''
  arXiv:hep-th/0503217.
}
\lref\mms{
  J.~M.~Maldacena, G.~W.~Moore and A.~Strominger,
  ``Counting BPS black holes in toroidal type II string theory,''
  arXiv:hep-th/9903163.
}

\lref\gv{
  R.~Gopakumar and C.~Vafa,
  ``M-theory and topological strings. II,''
  arXiv:hep-th/9812127.
}

\lref\DijkgraafXW{
  R.~Dijkgraaf, G.~W.~Moore, E.~Verlinde and H.~Verlinde,
  ``Elliptic genera of symmetric products and second quantized strings,''
  Commun.\ Math.\ Phys.\  {\bf 185}, 197 (1997)
  [arXiv:hep-th/9608096].
}

\lref\borch{R. E. Borcherds, "Automorphic forms on
$O_{s+2,2}(R)$ and infinite products" Invent. Math. {\bf 120}
(1995) 161.}

\lref\KAWAI{
  T.~Kawai,
  ``$N=2$ heterotic string threshold correction, $K3$ surface and generalized
  Kac-Moody superalgebra,''
  Phys.\ Lett.\ B {\bf 372}, 59 (1996)
  [arXiv:hep-th/9512046].
}

\lref\DVV{
  R.~Dijkgraaf, E.~Verlinde and H.~Verlinde,
  ``Counting dyons in N = 4 string theory,''
  Nucl.\ Phys.\ B {\bf 484}, 543 (1997)
  [arXiv:hep-th/9607026].
}

\lref\gava{
  I.~Antoniadis, E.~Gava, K.~S.~Narain and T.~R.~Taylor,
  ``N=2 type II heterotic duality and higher derivative F terms,''
  Nucl.\ Phys.\ B {\bf 455}, 109 (1995)
  [arXiv:hep-th/9507115].
}

\lref\fk{
  S.~Ferrara and R.~Kallosh,
  ``Universality of Supersymmetric Attractors,''
  Phys.\ Rev.\ D {\bf 54}, 1525 (1996)
  [arXiv:hep-th/9603090].
}

\lref\moorecharges{
  R.~Minasian and G.~W.~Moore,
  ``K-theory and Ramond-Ramond charge,''
  JHEP {\bf 9711}, 002 (1997)
  [arXiv:hep-th/9710230].
}

\lref\BMPV{
  J.~C.~Breckenridge, R.~C.~Myers, A.~W.~Peet and C.~Vafa,
  ``D-branes and spinning black holes,''
  Phys.\ Lett.\ B {\bf 391}, 93 (1997)
  [arXiv:hep-th/9602065].
}

\lref\adstop{
  D.~Gaiotto, A.~Strominger and X.~Yin,
  ``From AdS(3)/CFT(2) to black holes / topological strings,''
  arXiv:hep-th/0602046.
}

\lref\ascv{ A.~Strominger and C.~Vafa,
  ``Microscopic Origin of the Bekenstein-Hawking Entropy,''
  Phys.\ Lett.\ B {\bf 379}, 99 (1996)
  [arXiv:hep-th/9601029].}

\lref\iqbalnote{ A. Iqbal, ``A note on E-strings",
[arXiv:hep-th/0206064].}

\lref\ssyii{
  D.~Shih, A.~Strominger and X.~Yin,
  ``Counting dyons in N = 8 string theory,''
  arXiv:hep-th/0506151.
}

\lref\ddmpi{A.~Dabholkar, F.~Denef, G.~W.~Moore and B.~Pioline,
  ``Exact and asymptotic degeneracies of small black holes,''
  [arXiv:hep-th/0502157].
}

\lref\ddmp{
  A.~Dabholkar, F.~Denef, G.~W.~Moore and B.~Pioline,
  ``Precision counting of small black holes,''
  JHEP {\bf 0510}, 096 (2005)
  [arXiv:hep-th/0507014].
}

\lref\cardosoii{G.~L.~Cardoso, B.~de Wit, J.~Kappeli and T.~Mohaupt,
  ``Asymptotic degeneracy of dyonic N = 4 string states and black hole
  entropy,''
  JHEP {\bf 0412}, 075 (2004)
  [arXiv:hep-th/0412287].
}

\lref\VafaQA{
  C.~Vafa,
  ``Two dimensional Yang-Mills, black holes and topological strings,''
  arXiv:hep-th/0406058.
}

\lref\qdeform{
  M.~Aganagic, H.~Ooguri, N.~Saulina and C.~Vafa,
  ``Black holes, q-deformed 2d Yang-Mills, and non-perturbative topological
  Nucl.\ Phys.\ B {\bf 715}, 304 (2005)
  [arXiv:hep-th/0411280].
}

\lref\neitzke{
  M.~Aganagic, A.~Neitzke and C.~Vafa,
  ``BPS microstates and the open topological string wave function,''
  arXiv:hep-th/0504054.
}

\lref\babyu{
  R.~Dijkgraaf, R.~Gopakumar, H.~Ooguri and C.~Vafa,
  ``Baby universes in string theory,''
  Phys.\ Rev.\ D {\bf 73}, 066002 (2006)
  [arXiv:hep-th/0504221].
}

\lref\DVafaV{
  R.~Dijkgraaf, C.~Vafa and E.~Verlinde,
  ``M-theory and a topological string duality,''
  arXiv:hep-th/0602087.
}

\lref\verlindeanomaly{
  E.~P.~Verlinde,
  ``Attractors and the holomorphic anomaly,''
  arXiv:hep-th/0412139.
}

\lref\largenequalsfour{
  S.~Gukov, E.~Martinec, G.~W.~Moore and A.~Strominger,
  ``An index for 2D field theories with large N = 4 superconformal  symmetry,''
  arXiv:hep-th/0404023.
}

\lref\deboerlarge{
  J.~de Boer, A.~Pasquinucci and K.~Skenderis,
  ``AdS/CFT dualities involving large 2d N = 4 superconformal symmetry,''
  Adv.\ Theor.\ Math.\ Phys.\  {\bf 3}, 577 (1999)
  [arXiv:hep-th/9904073].
}

\lref\deboer{
  J.~de Boer,
  ``Six-dimensional supergravity on S**3 x AdS(3) and 2d conformal field
  Nucl.\ Phys.\ B {\bf 548}, 139 (1999)
  [arXiv:hep-th/9806104].
}

\lref\denefwork{
  F.~Denef and G.~W.~Moore, to appear and talk given at Strings 2006.
}

\lref\verlindework{ J.~de Boer, M.~Cheng, R.~Dijkgraaf, J.~Manschot
and E.~Verlinde, to appear and talk given at Strings 2006. }

\lref\freedwitten{
  D.~S.~Freed and E.~Witten,
  ``Anomalies in string theory with D-branes,''
  arXiv:hep-th/9907189.
}

\lref\minasianmoore{
  R.~Minasian, G.~W.~Moore and D.~Tsimpis,
  Commun.\ Math.\ Phys.\  {\bf 209}, 325 (2000)
  [arXiv:hep-th/9904217].
}

\lref\ourquinticpaper{
  D.~Gaiotto, M.~Guica, L.~Huang, A.~Simons, A.~Strominger and X.~Yin,
  ``D4-D0 branes on the quintic,''
  JHEP {\bf 0603}, 019 (2006)
  [arXiv:hep-th/0509168].
}

\lref\moorebelov{
  D.~Belov and G.~W.~Moore,
  ``Holographic action for the self-dual field,''
  arXiv:hep-th/0605038.
}

\lref\fareytail{ R.~Dijkgraaf, J.~M.~Maldacena, G.~W.~Moore and
E.~P.~Verlinde,
  ``A black hole farey tail,''
  arXiv:hep-th/0005003.
}

\lref\marcus{
  A.~Klemm and M.~Marino,
  ``Counting BPS states on the Enriques Calabi-Yau,''
  arXiv:hep-th/0512227.
}

\lref\toroidal{
  J.~M.~Maldacena, G.~W.~Moore and A.~Strominger,
  ``Counting BPS black holes in toroidal type II string theory,''
  arXiv:hep-th/9903163.
}



\Title{\vbox{\baselineskip12pt\hbox{} }} {\vbox{\centerline{The
M5-Brane Elliptic Genus:}\bigskip \centerline{Modularity and BPS
States}}}

\centerline{Davide Gaiotto,~ Andrew Strominger and Xi Yin }
\smallskip
\centerline{Jefferson Physical Laboratory, Harvard University,
Cambridge, MA 02138} \vskip .6in \centerline{\bf Abstract} { The
modified elliptic genus for an M5-brane wrapped on a four-cycle of a
Calabi-Yau threefold encodes the degeneracies of an infinite set of
BPS states in four dimensions. By holomorphy and modular invariance,
it can be determined completely from the knowledge of a finite set
of such BPS states. We show the feasibility of such a computation
and determine the exact modified elliptic genus for an M5-brane
wrapping a hyperplane section of the quintic threefold.
 } \vskip .3in

\Date{July 2006}

\listtoc\writetoc

\newsec{Introduction}

An exact counting of BPS states in ${\cal N}=2, d=4$
compactifications of (say type IIA) string theory is a difficult
problem. For large charges these objects have dual descriptions as
supersymmetric black holes.  The leading and subleading order
entropies of such black holes in M-theory has been explained using
the $(0,4)$ CFT on the effective string coming from a wrapped
M5-brane \msw. Herein we refine the analysis of \msw\ and consider
the modified elliptic genus (to be defined later) which counts BPS
states of D4-D2-D0 system on a generic Calabi-Yau threefold. As
emphasized in
\refs{\fareytail,\moorebelov,\verlindework,\denefwork}, modular
invariance imposes strong constraints on the elliptic genus, and
determines it completely in terms of a finite number of coefficients
in its $q$-expansion. In this paper we will use this fact to derive
the exact partition function of a basic class of BPS states in type
IIA string compactified on the quintic threefold. Our results shed
light on the conjectured relation\refs{\osv\dab\ddmpi\ddmp\marcus
\VafaQA\qdeform\neitzke\verlindeanomaly\babyu\adstop-\DVafaV}
between the exact partition function of the black hole and the
topological string amplitude.

An alternative approach - related by M/IIA duality - to count the
degeneracy of D4-D2-D0 bound states is to quantize the classical
moduli space of supersymmetric D-brane configurations. The latter
involve wrapped D4-brane with fluxes bound to pointlike instantons.
The possible $U(1)$ fluxes on the D4-brane are in 1-1 correspondence
with divisors (an algebraic sum of holomorphic curves) on its world
volume, and in particular involves nontrivial information of
Gromov-Witten invariants. We use this method to compute the
coefficients of the modified elliptic genus in the example of the
quintic for small charges corresponding to the first few terms in
the $q$-expansion. A priori, one might expect the results to
overdetermine the elliptic genus: in principle, a few of these
numbers are sufficient to determine the elliptic genus, and all
other coefficients are predicted based on the modular property. This
would give a sharp test of the direct counting of supersymmetric
D4-D2-D0 bound states. In the simple example we consider, the system
of minimal D4-brane on the quintic, this agreement requires highly
nontrivial relations among Gromov-Witten invariants of various
degrees. Miraculously, we find this relation to hold in all the
coefficients we computed based on the geometry of the classical
moduli space, up to a small ambiguity due to singularities in the
moduli space, corresponding to colliding pointlike D0 instantons.

We propose an approach to resolve this ambiguity, based on the
relevant dual M-theory $AdS_3\times S^2\times CY$ attractor geometry
\adstop. In this picture the needed coefficients of the
$q$-expansion are supplied by the degeneracies of a few low-lying
BPS states of gravitons and wrapped (anti-)M2-branes. It is not
clear to us why this dual picture should be valid, since some of the
charges are small in all of our our examples. Nevertheless, with
this approach, we find perfect agreement of the degeneracy of BPS
states with the relations expected from the modular property of the
elliptic genus. (This in turns suggests that there should be some
justification for our approach.)

This approach sheds light on the conjecture \osv\ that the
Gromov-Witten invariants are captured by the black hole partition
function. Here we see that the $U(1)$ fluxes on the D4-brane, which
are related to curves in the CY subject to the constraint that it
must lie on the world volume of the D4-brane, carry information
about Gromov-Witten invariants. This extra constraint becomes
unimportant for large D4-brane charges (a sufficiently high degree
hypersurface can be made to pass through any given collection of
curves), in which limit the (square of the) topological string
partition function becomes a good approximation of the black hole
partition function.

In section 2, we describe the general structure of the modified elliptic genus
of the MSW $(0,4)$ CFT. We argue that the elliptic genus has simple
anti-holomorphic dependence, and can be determined by a finite number
of holomorphic characters that transform in a known representation of
$SL(2,{\bf Z})$. In section 3, we study the $(0,4)$ CFT for an M5-brane wrapped
on the hyperplane section in the quintic. We count the degeneracy of a number of
BPS states based on the classical moduli space of D-brane configurations,
as well as a hypothetical chiral ring structure motivated by a dilute gas
approximation in the $AdS_3$ dual. These results are compared to the
modular property of the elliptic genus, and surprising agreements are found.
In particular, we conjecture an exact expression for the elliptic genus
in this case. Section 4 studies a different example, a free ${\bf Z}_5$ quotient of
the Fermat quintic. In this case we present the structure of the elliptic genus as determined
by its modular property, although the direct counting based on the classical
moduli space is more difficult than the quintic, and is left to future work.
We conclude in section 5.

Results related to those of section 2 on the structure of the
modified elliptic genus have been independently obtained by Denef
and Moore \denefwork, and de Boer, Cheng, Dijkgraaf, Manschot and E.
Verlinde \verlindework .

\newsec{The M5-brane $(0,4)$ CFT}

We will be considering an M-theory 5-brane wrapped on a 4-cycle $P$ in Calabi-Yau
space $X$, and extended in ${\bf R}^{1,4}$. The low energy excitations of the
M5-brane can be described by an effective $1+1$ dimensional CFT \msw, which has
$(0,4)$ superconformal symmetry. If one further compactifies the direction in which
the M5-brane extends in ${\bf R}^{1,4}$ on a circle, one obtains a wrapped D4-brane
in type IIA string theory compactified on $X$.
Excitations of the M5-brane induce M2-brane charges
and can carry momenta. These in general correspond to D4-D2-D0 bound states
in type IIA string theory. The
correspondence between M5-brane states and D4-D2-D0 bound states
is understood and will be used freely in this paper.

In the following, we will follow the convention of \msw\ and denote
by $6D_{ABC}$ the intersection numbers in a basis $\Sigma_A$ of
$H_4(X,{\bf Z})$. The D4, D2, D0 charges will often be labelled
$p^A, q_A, q_0$, respectively. $D_{AB}\equiv D_{ABC}p^C$, $D\equiv
D_{ABC}p^Ap^Bp^C$, and $D^{AB}$ is the inverse matrix of $D_{AB}$.
The attractor K\"ahler class of $X$ is proportional to
$J=p^A\omega_A$ where $\omega_A$ is a basis of harmonic $(1,1)$
forms dual to $\Sigma_A$.

\subsec{General structure}

The M5-brane $(0,4)$ CFT \msw\ has central
charge $c_L=6D+c_2\cdot P$, $c_R=6D+\half c_2\cdot P$. There are 3
noncompact bosons $X^i$ on the left and right, corresponding to the
collective coordinates in the transverse ${\bf R}^3$. There are free
bosons $\phi_A$, coming from the world volume anti-symmetric tensor
field reduced on $\omega_A$. Namely
\eqn\tphi{ T \sim D^{AB} d\phi_A \wedge \omega_B }
The self-duality of $T$ implies that $\varphi=p^A\phi_A$ is purely right moving.
Together with four goldstinos $\tilde\psi^{\pm\pm}$, corresponding to
the four supersymmetries broken by the M5-brane, $(\bar\partial
X^i, \bar\partial \varphi, \tilde\psi^{\pm\pm})$ form a right moving
${\cal N}=4$ multiplet. The $\phi_A$'s are compactified on a lattice
of signature $(h^{1,1}(X)-1,1)$,
with the intersection pairing given by $-{1\over 6}D^{AB}$, where
the $(h^{1,1}-1)$ bosons orthogonal to $\varphi$ are left movers.

This CFT in fact has ${\cal A}_{k^+,\infty}$ symmetry algebra
($k^+={c_R\over 6}$), which is a Wigner contraction of the large
${\cal N}=4$ superconformal algebra. Writing $U=\bar\partial
\varphi$, in addition to the small ${\cal N}=4$ SCA relations, there
are OPEs \eqn\opsdef{ \eqalign{ & \bar G^{\alpha a}(\bar z) U(0)
\sim {\tilde \psi^{\alpha a}(0)\over \bar z} \cr & \bar G^{\alpha
a}(\bar z) \tilde\psi^{\beta b}(0) \sim
{\epsilon^{\alpha\beta}\epsilon^{ab} U(0) \over \bar z} \cr &
J_R^i(\bar z) \tilde\psi^{\alpha a}(0) \sim
{{(\sigma^i)^\alpha}_\beta \tilde\psi^{\beta a}(0)\over \bar z} } }

We will restrict to the case where $X$ is a Calabi-Yau manifold of
full $SU(3)$ holonomy. Let $\Lambda = H^2(P,{\bf Z})$ be the
cohomology lattice of $P$. In general $P$ may not be a spin
manifold, and $\Lambda$, which is self-dual, may not be even. In
this case there is Freed-Witten anomaly \freedwitten, which requires
one to turn on a half-integral flux $\half c_1(P)$ in addition to
integral fluxes on the M5-brane world volume. We will write
$J=p^A\omega_A$ to represent the class of $P$ itself, or the
cohomology class dual to $P\cap P$ in $H_2(P)$, which is the same as
$-c_1(P)$. One can write a modular invariant theta function
associated with the lattice $\Lambda$, \eqn\modthe{
\Theta_\Lambda(\tau,\bar\tau, y) = \sum_{v\in\Lambda+\half J}
(-)^{p\cdot q(v)} e^{-\pi i \tau v_-^2-\pi i\bar\tau v_+^2+2\pi i
q(v)\cdot y} } Here $v_+$ and $v_-$ are the self-dual and
anti-self-dual projections of lattice vector $v$, or equivalently,
projections along $J$ and $J^\perp$. $v_\pm^2$ are defined using the
intersection form on $\Lambda$. $q(v)$ is the natural projection of
$v$ from $H_2(P,{\bf Z})$ to $H_2(X,{\bf Z})$, corresponding to the
M2/D2-brane charge. Note that $J$ is an characteristic element of
$H^2(P,{\bf Z})$, hence $v^2+q(v)\cdot p\equiv 0$ mod 2,
$(-)^{p\cdot q(v)}=1$ when $\Lambda$ is even.

$\Lambda_X=H^2(X,{\bf Z})$ is embedded in $\Lambda$ as a sublattice.
It consists of charge vectors with $q_A=6D_{AB}k^B$, $k^A\in {\bf
Z}$. $\Lambda_X^\perp\subset \Lambda$ consists of only left-moving
lattice vectors. The theta function \modthe\ can be decomposed as
\eqn\absb{ \Theta_\Lambda(\tau,\bar\tau,y) = \sum_{\delta}
\Theta_{\Lambda_X^\perp+\delta}(\tau)
\Theta_{\Lambda_X+\delta}(\tau,\bar\tau, y) } where $\delta$ runs
through a finite set of $\det(6D_{AB})$ shift vectors. The functions
$\Theta_{\Lambda_X+\delta}$, which we will abbreviate as
$\Theta_{\delta}$, can be written explicitly \eqn\tehtaa{\eqalign{
\Theta_\delta(\tau,\bar\tau, y) = \sum_{q_A=6D_{AB}(k^B+\half
p^B)+\delta_A,~k^A\in{\bf Z}} (-)^{p^Aq_A} \exp\left[{{2\pi
i\tau\over 12}\left( {p^Ap^B\over D}-D^{AB} \right) q_A q_B}\right.
\cr \left. - {2\pi i\bar\tau\over 12 D}(p^A q_A)^2 +2\pi iy^Aq_A
\right] } }

\subsec{The modified elliptic genus}

We are interested in computing a supersymmetric index of the $(0,4)$
CFT. The elliptic genus vanishes, due to the degeneracy of Ramond
ground state generated by the zero modes of the goldstinos
$\psi_0^{\pm\pm}$. As explained in the previous subsection, the CFT
has symmetry algebra ${\cal A}_{k^+,\infty}$, which extends the
small ${\cal N}=4$ superconformal algebra. One can define a modified
elliptic genus \toroidal, \eqn\modff{ Z(\tau,\bar\tau,y) = {\rm
Tr}_R \half F^2(-)^{F} q^{L_0-{c_L\over 24}} \bar q^{\bar
L_0-{c_R\over 24}} e^{2\pi iy^A Q_A} } where $q=e^{2\pi i\tau}$,
$Q_A$ are the charges associated with the free bosons $\phi_A$,
corresponding to induced M2-brane charges on the M5-brane. $F$ is a
fermion number, which can be identified with $2J_R^3+p^AQ_A$, where
$J_R$ is the right moving R-charge, and $p^AQ_A$ is the contribution
from the quantization of the self-dual 3-form field (this is an
example of a general phenomenon explained in \moorebelov).

An important point is that not only the right moving ground states
contribute to $Z$. Let $|0\rangle$ be a ground state, then $\tilde
\psi_0^{\pm\pm}$ acting on $|0\rangle$ generates a multiplet of 4
states, contributing $1$ to ${\rm Tr}{F^2\over 2}(-)^F$. These
states are annihilated by the supercharges $\bar G_0^{\pm\pm}$. Now
consider a state of charge $q_A$ under the current $j_A=d\phi_A$,
say $|q\rangle=e^{iD^{AB}q_A\phi_B}|0\rangle$. $\bar G_0^{\pm\pm}$
acting on $p^A\phi_A$ gives rise to its fermionic partners $\tilde
\psi_0^{\pm\pm}$, and leave the left moving fields invariant. Hence
\eqn\sfog{ (\bar G_0^{\pm\pm}-p^AQ_A\tilde
\psi_0^{\pm\pm})|q\rangle=0 } This is simply saying that the state
$|q\rangle$ preserves supersymmetries nonlinearly, and acting with
$\bar G_0^{\pm\pm}$ doesn't give rise to new states other than the
multiplet generated by $\tilde\psi_0^{\pm\pm}$. This is in accord
with the fact that $q_A$ correspond to induced D2-brane charges, and
D4-D2 bound states preserve different sets of supersymmetries than
that of D4(-D0). Therefore this multiplet also contributes $1$ to
the modified elliptic genus. In general the modified elliptic genus
$Z$ receives contribution from states of charge $q_A$ with
\eqn\lobar{(\bar L_0-{c_R\over 24})|\psi\rangle = {(p^AQ_A)^2\over
12D}|\psi\rangle}

As in \refs{\ddmp, \denefwork, \verlindework}, the modified elliptic
genus has the general form \eqn\zzzfom{ Z(\tau,\bar\tau, y) =
\sum_\delta Z_\delta(\tau) \Theta_\delta(\tau,\bar\tau,y) } where
$\Theta_\delta$ are given by \tehtaa. This structure can be argued
using the fact that shifting $B$-field by an integral amount on $X$
does not change the degeneracy of D4-D2-D0 bound states, but
generates additional D2 and D0-brane charges, corresponding to a
translation in $\Lambda_X$. Another way to think about \zzzfom\ is
that the theta function of the cohomology lattice vectors of
$H^2(P,{\bf Z})$ that do not correspond to conserved charges get
completed into holomorphic characters $Z_\delta(\tau)$ in the full
CFT.

$\Theta_\delta(\tau,\bar\tau, y)$ form a modular representation in terms of
weight $(\half(h^{1,1}(X)-1),\half)$ Jacobi forms. The $T$
transformation is represented by the matrix \eqn\tmatas{
T^\Theta_{\delta\lambda} = \delta_{\delta\lambda} \exp\left( -{2\pi
i\over 12}D^{AB}\delta_A \delta_B \right) } The $S$ transformation
is represented by \eqn\smatss{ S^\Theta_{\delta\lambda} = {1\over
\sqrt{6D}} \exp\left( -{2\pi i\over 6}D^{AB}\delta_A \lambda_B
\right) }

The modified elliptic genus of the MSW $(0,4)$ CFT is expected to be
a weight $(-{3\over 2},{1\over 2})$ Jacobi form. The left weight
$-{3\over 2}$ comes from the three noncompact bosons, and the right
weight is modified by the insertion of $F^2$ to $-{3\over2}+2=\half$.
$Z_{\delta}(\tau)$
transform under $SL(2,{\bf Z})$ with $T=(T^{\Theta})^*$,
$S=(S^\Theta)^*$ up to an overall phase that is easy to determine.

Knowing the modular representation of $Z_\delta(\tau)$, one can
determine all of them from the polar terms in the $q$-expansion of
$Z_\delta$, where $q=e^{2\pi i\tau}$, via the generalized Rademacher
expansion. Equivalently, there is a basis of modular vectors transforming
the same way as $Z_\delta(\tau)$ with the most singular polar term
$q^{-{c_L\over 24}}$,
whose number is the same as the number
of possible polar terms of $Z_\delta(\tau)$.
 We will explore the constraints of modular invariance of
$Z$ on the degeneracy of BPS states in the rest of this paper.

\newsec{BPS states on the quintic}

In this section $X$ will be the quintic 3-fold with a generic
complex structure. We will study the $(0,4)$ CFT associated with an
M5-brane wrapped on the hyperplane section $P$ in $X$. $J$ will
refer to the hyperplane class. This CFT has $c_L=55, c_R=30$, and
$D={5\over 6}$.

\subsec{A naive counting from geometry}

Our strategy will be to count the D4-D2-D0 bound states of given
charges $(p=1, q_1, q_0)$ by computing the Euler character of the
classical moduli space of the branes. The D4-brane is wrapped on the
hyperplane section $P$, and it is free to move in its moduli space
${\bf P}^4$. In a supersymmetric configuration, the D2-branes are
dissolved into fluxes on the D4 world volume. We will assume that
D0-branes can be either pointlike instantons (which can form bound
states among themselves) on the D4-brane world volume, or dissolve
into smooth $U(1)$ fluxes.

One may attempt to describe the moduli space of classically
supersymmetric D4-D2-D0 configuration as a fibration over the D4
moduli space, the fiber being the Hilbert scheme of points on the
D4-brane world volume etc. This is a useful approximation in the
limit of large D0-brane charges, but is difficult to apply for small charges.
The reason is that the D4-brane world volume degenerates in various
loci in its moduli space. It turns out that for the cases we will be
computing, it is more useful to describe the moduli space by first
fixing the D0-branes, and consider the space of D4-branes that pass
through these D0-branes and admit certain classes of fluxes.

Due to Freed-Witten anomaly one must turn on half integral flux, say
$F={1\over 2}J$, on the D4-brane. We will call this the ``pure"
D4-brane. There is induced D2-brane charge ${5\over 2}$ and D0-brane
charge \refs{\moorecharges,\ourquinticpaper} $$ -{\chi(P)\over
24}-\half\int_P F\wedge F=-{35\over 12} $$ In the CFT this
corresponds to a state with $L_0=0$ and $(\bar L_0-{c_R\over 24}) =
{(p^Aq_A)^2\over 12 D}={5\over 8}$. In the following we will label
states by their additional D2-brane charge $\Delta q_1$, as well as
the additional D0-brane charge $\Delta q_0$.\foot{$\Delta q_0$ is
related to $L_0$ and $\bar L_0$ by $\Delta q_0-{35\over
12}=(L_0-{c_L\over 24})- (\bar L_0-{c_R\over 24})=(L_0-{55\over
24})-{1\over 10}(\Delta q_1+{5\over 2})^2$.}

\bigskip

\noindent $\bullet$ $\Delta q_1=0,\Delta q_0=0~(L_0=0)$

This is the ``pure" D4-brane, which is free to move around its
moduli space ${\bf P}^4$. The number of supersymmetric states is
$\chi({\bf P}^4)=5$.

\bigskip

\noindent $\bullet$ $\Delta q_1=0,\Delta q_0=1~(L_0=1)$

Next we consider the D4 bound to a single D0-brane. Requiring the
D4-brane to pass through the D0, the moduli space of D4-D0 is ${\bf
P}^3$ fibered over $X$. It has Euler character $\chi({\bf
P}^3)\chi(X)=-800$.

\bigskip

\noindent $\bullet$ $\Delta q_1=0,\Delta q_0=2~(L_0=2)$

Let us consider the D4 bound to 2 D0's. The two D0-branes can either
both be free to move on the D4-brane world volume, or bind together
as a single pointlike object. The latter is counted the same way as
the $\Delta q_0=1$ case, whereas the former is described by the
moduli space as ${\bf P}^2$ fibered over ${\rm Sym}^2(X)$. We shall
ignore the subtle contribution from the locus in the moduli space
where the two D0's coincide. The number of states, counted with
sign, is then $\chi({\bf P}^2)\chi({\rm Sym}^2(X))+\chi({\bf
P}^3)\chi(X)=58900$.

\bigskip

\noindent $\bullet$ $\Delta q_1=0,\Delta q_0=3~(L_0=3)$

The states with $\Delta q_0=3$ involves the D4 bound to 3 pointlike
instantons, as well as a D4-brane with flux $F=C_1-C_1'$, where
$C_1$ and $C_1'$ are two different degree 1 rational curves in $X$
that lie in $P$, we write them for their dual harmonic forms on $P$.
The latter can happen only when $P$ passes through both $C_1$ and
$C_1'$. $C_1$ and $C_1'$ generically do not touch, and each have
self-intersection number $C_1\cdot C_1=-3$.\foot{This follows from
the adjunction formula, for a curve $C$ in $P$, $C\cdot C+C\cdot
J=2g-2$.} The flux $F$ gives rise to induced D0-brane charge $-\half
F^2 = 3$. Generically this condition fixes a unique hyperplane, and
we have $2875\times 2874$ choices of $C_1-C_1'$. These give rise to
(again, ignoring the subtlety where the D0's coincide in the moduli
space)
$$
\chi({\bf P}^1)\chi({\rm Sym}^3(X))+\chi({\bf P}^2)
\chi(X)^2+\chi({\bf P}^3)\chi(X)+2875\cdot 2874 = 5755150
$$
states.

\bigskip

\noindent $\bullet$ $\Delta q_1=1,\Delta q_0=1~(L_0={8\over 5})$

Next we consider the D4 bound to a single D2-brane. This can be
realized by turning on a flux $F=C_1$ on the D4-brane world volume (in
addition to the original ${J\over 2}$),
where $C_1$ is (dual to) a degree 1 rational curve. It gives rise to
additional induced D0-brane charge $-\half (F+{J\over 2})^2
+{5\over 8}=1$. Requiring the D4-brane to pass
through the curve $C_1$ reduces its moduli space from ${\bf P}^4$ to
${\bf P}^2$. There are 2875 degree 1 rational curves $C_1$. Our naive
quantization of the classical moduli space doesn not determine the
overall fermion number of these states. However, we can fix the fermion
number by comparison with the holographic dual, and these states turn out
to have an odd fermion number.\foot{
A D4-brane with flux
$F=C_1$ corresponds to an M2-brane wrapped on $C_1\subset X$ in
the dual $AdS_3\times S^2\times X$ geometry. The chiral primary
state associated with this wrapped M2-brane is a fermion.}
This description will be explored in the next section.
In the end, we get the counting $\chi({\bf
P}^2)\cdot (-2875)=-8625$.

Similarly, the same degeneracy applies to the states with $\Delta
q_1=-1, \Delta q_0=2~(L_0={8\over 5})$.

\bigskip

\noindent $\bullet$ $\Delta q_1=1, \Delta q_0=2~(L_0={13\over 5})$

Such states involve a D4 with flux $F=C_1$ where $C_1$ is a degree 1
rational curve, as well as a pointlike D0 instanton. We don't
understand precisely what happens when the D0-brane coincides with
$C_1$, and will again ignore this subtlety for now. When the D0-brane is away
from $C_1$, the D4-brane that passes through both the D0-brane and the curve $C_1$
has moduli space ${\bf P}^1$. The number of states counted with sign
is
$$
(-2875)\cdot \chi(X) \chi({\bf P}^1) =1150000.
$$

\bigskip

\noindent $\bullet$ $\Delta q_1=2, \Delta q_0=1~(L_0={12\over 5})$

The states with D2-brane charge $\Delta q_1=2$ has a minimal
D0-brane charge 1, coming from the D4-brane with flux $F=C_2$, where
$C_2$ is a degree 2 rational curve. There are 609250 such curves.
Hyperplanes that pass though a degree 2 curve have moduli space
${\bf P}^1$. We have degeneracy\foot{The sign, once again, is
determined by comparison to the $AdS_3$ dual.}
$(-609250)\cdot\chi({\bf P}^1)=-1218500$.

\bigskip

\noindent $\bullet$ $\Delta q_1=2, \Delta q_0=2~(L_0={17\over 5})$

The counting of such states receives three kinds of contributions: a
D4 with flux $F=C_1+C_1'$ where $C_{1},C_1'$ are degree 1 rational
curves; D4 with flux $F=C_2$ where $C_2$ is a degree 2 rational
curve, bound to a D0 pointlike instanton; and D4 with flux $F=J-C_3$
where $C_3$ is a degree 3 rational curve. The counting is (again,
ignoring the subtle case when the D0 coincides with $C_2$)
$$
\half \,2875\cdot 2874 + 2875 \cdot \chi({\bf P}^2) + (-609250)\cdot
\chi(X) +317206375 = 443196375.
$$
One might think that there is another contribution, coming from D4
with flux $F=J-C_3'$ bound to a single D0, where $C_3'$ is a degree
3 {\sl genus 1} curve. These however give rise to the same set of
fluxes as $F=C_2$.\foot{To see whether $F_1=C_2$ and $F_2=J-C_3'$
are the same flux on $P$, one simply needs to check whether
$(F_1^--F_2^-)^2=0$, where $F^-$ is the anti-self-dual projection of
$F$, $F^-=F-{1\over 5}(F\cdot J)J$. This is the case if and only if
$C_2\cdot C_3'=6$. In fact, any $C_3'$ can be defined by the
equations of the form $P_3(x^i)=0, H_1(x^i)=G_1(x^i)=0$, where $P_3$
is a cubic polynomial in homogeneous coordinates $x^i$ on ${\bf
P}^4$, and $H_1, G_1$ are linear polynomials. One can take $H_1$ to
be the hyperplane section $P$. The quintic equation must be of the
form $P_3(x^i)P_2(x^i)+H_1(x^i) Q_4(x^i)+G_1(x^i)R_4(x^i)=0$. Now
$P_2(x^i)$ together with $H_1, G_1$ define a degree 2 rational curve
$C_2$, which touches $C_3'$ at 6 points. Indeed, there are 609250
degree 3 genus 1 curves in the quintic, the same as the number of
degree 2 genus 0 curves. }

\subsec{Connection to topological strings}

The M5-brane $(0,4)$ CFT is dual to M-theory on $AdS_3\times
S^2\times X$ attractor geometry \refs{\deboer, \adstop}. It was
shown in \adstop\ that the elliptic genus that counts the chiral
primaries coming from supergravity modes as well as CY-wrapped M2
and anti-M2 branes in the dilute gas approximation reproduces the
square of the topological string partition function on $X$. The
supergravity picture is not expected to be generally valid for small
M5-brane charges/fluxes, although some quantities may be BPS
protected. However there appears to be a rough correspondence
between supersymmetric ground states of the D4-brane bound to D2,
D0-branes and multi-particle chiral primaries in $AdS_3$:
$$
\eqalign{ {\rm pointlike~D0~instantons~}& \longleftrightarrow {\rm~
massless~supergravity~modes} \cr {\rm flux~~}F=\sum_i C_i-\sum_j C_j'~ &
\longleftrightarrow ~{\rm M2~wrapped~on}~C_i,~~~{\rm
\overline{M2}~wrapped~on~}C_j'
 }
$$
where $C_i, C_j'$ are holomorphic curves in $X$.

Under the spectral flow of ${\cal A}_{k^+,\infty}$ algebra from NS
to Ramond sector, which includes $L_0\to L_0 - J_R^3+{c_R\over 24}$
and a shift of the membrane charges $Q_A\to Q_A+3D_{AB}p^B$ (see
also \verlindework), the chiral primaries flow to Ramond sector
states of the form \sfog, \lobar. The unitarity bound on the chiral
primaries takes the form $\bar L_0 = J_R^3 + {(p^A Q_A)^2\over
12D}$. In particular, since $(J_R)^-_1$ flows to $(J_R)^-_0$, the
chiral primaries are annihilated by $(J_R)^-_1$, and flow to lowest
$SU(2)_R$ weight states in the Ramond sector. For example, the
$AdS_3$ vacuum flows to a lowest weight state of spin $J_R^3 =
-{c_R\over 12}$. Together with the states obtained by acting with
$\tilde \psi^{+\pm}$, this $SU(2)_R$ multiplet contributes to the
modified elliptic genus with degeneracy $2(j_R-\half)+1 ={1\over
6}c_R = D+{1\over 12} c_2\cdot P$ (with the insertion of $\half F^2$
in \modff\ absorbed). This is precisely the Euler character of the
moduli space of a ``pure" D4-brane of charge $p^A$, namely ${\bf
P}^{D+{1\over 12} c_2\cdot P-1}$.

The prescription to compute the elliptic genus is \eqn\ellshg{
Z(\tau,\bar\tau,y)={\rm Tr}_{ch.pr.} (-)^F ({c_R\over 6}-2J_R^3)
q^{L_0-{c_L\over 24}}\bar q^{{(p\cdot Q')^2\over 12D}} e^{2\pi i y^A
Q_A'} } where $Q_A'\equiv Q_A+3D_{AB}p^B$, and $({c_R\over
6}-2J_R^3)$ is the contribution due to the $SU(2)_R$ multiplets
described above.

We propose to compute the first few terms in the elliptic genus
using the dilute gas approximation in the $AdS_3$. This involves
a free gas of massless supergravity modes
and wrapped M2 and anti-M2-branes. They can carry
angular momenta on the $S^2$ as well as in the
$AdS_3$.
In this approach, one can determine the fermion number of the chiral
primaries corresponding to the wrapped M2-branes, as in \adstop, which
is hard to determine by directly quantizing the classical moduli space of D4-branes.

Let us denote by ${\cal O}_{n,j}^a$ the chiral primary operator dual
to a graviton/hyper/vector multiplet of spin $j$ on the $S^2$, and
$L_0-\bar L_0=n$; and ${\cal O}^C_{n,j}$ (${\cal O}^{-C}_{n,j}$) the
chiral primaries dual to M2-brane (anti-M2-brane) carrying spin $j$
and $L_0-\bar L_0=n$.\foot{ It would be useful to justify the chiral
ring generators and (lack of) relations directly from the sigma
model description of the $(0,4)$ CFT \minasianmoore. } The $L_{-1}$
descendants of the chiral primaries are dual to holomorphic
derivatives of the corresponding operators. We have $n=\half$ for
the massless hypermultiplets, $n=-1,0,1,2$ for the graviton
multiplet, and $n=0,1$ for the vector multiplets. And $n=\half$ for
an (anti-)M2-brane wrapped on a rational curve \adstop. In the case
of the quintic, the contribution to the elliptic genus from 204
hypermultiplets, 1 vector multiplet and 1 graviton multiplet is
equivalent to 200 hypermultiplets.

In the following we redo the counting in the previous subsection
using this dilute gas approximation (in the chiral ring language),
which does not have the ambiguity with singularities of the
classical moduli space of the D4-brane.

\bigskip

\noindent $\bullet$ $q_1=0, L_0=1$

There are 200 ${\cal O}_{\half,\half}^a$'s (counted with sign) that
contribute, each giving rise to a fermionic Ramond state of spin
$j_R=2-\half={3\over 2}$. Hence the contribution to the (modified)
elliptic genus is $-4\times 200=-800$. This is the same answer as
the one we obtained from the ``naive" counting.

\bigskip

\noindent $\bullet$ $q_1=0, L_0 = 2$

The states that contribute are given by operators of the form ${\cal
O}^a_{\half,\half} {\cal O}_{\half,\half}^{b},~\partial {\cal
O}_{\half,\half}^a,~{\cal O}_{\half,{3\over 2}}^a$, of spin $1,\half$ and
${3\over 2}$ respectively. The counting is then
$$
3\cdot (\half 200\cdot 199)-4\cdot 200-2\cdot 200 = 58500
$$
Note that this differs slightly from the answer 58900 we obtained
from the ``naive" counting.

\bigskip

\noindent $\bullet$ $q_1=0, L_0=3$

The states that contribute are given by operators of the form ${\cal
O}_{\half,\half}^a {\cal O}_{\half,\half}^b {\cal O}_{\half,\half}^c$, ${\cal
O}_{\half,\half}^a {\cal O}_{\half,{3\over 2}}^b$, ${\cal O}_{\half,\half}^a
\partial {\cal O}_{\half,\half}^b$, ${\cal O}_{\half,{5\over 2}}^a$, $\partial {\cal
O}_{\half,{3\over 2}}^a$, $\partial^2 {\cal O}_{\half,\half}^a$, ${\cal
O}^{C_1}_{\half,1}{\cal O}^{-C_1'}_{\half,1}$. The
counting is
$$
-2\cdot (200\cdot 199\cdot 198/6) + (3+1)\cdot 200 \cdot 200
-(4+2+0)\cdot 200+ 2875^2=5797625
$$

\bigskip

\noindent $\bullet$ $q_1=1, L_0={3\over 2}+{q^2\over 10}={8\over 5}$

The operators that contribute are ${\cal O}^{C_1}_{\half,1}$, which
gives rise to $-3\times 2875=-8625$ states, the same as the naive
counting in the previous section.

\bigskip

\noindent $\bullet$ $q_1=1, L_0 = {13\over 5}$

The operators that contribute are ${\cal O}^a_{\half,{1\over 2}}{\cal
O}^{C_1}_{\half,1}$, ${\cal O}^{C_1}_{\half,2}$, $\partial
{\cal O}^{C_1}_{\half,1}$. The counting is
$$
2\cdot (-200)\cdot (-2875)+(3+1)\cdot (-2875)=1138500
$$

\bigskip

\noindent $\bullet$ $q_1=2, L_0 = 2+{q^2\over 10}={12\over 5}$

The operators that contribute are ${\cal O}^{C_2}_{\half,{3\over 2}}$. The
counting is the same as before, giving rise to $-1218500$ states.

\bigskip

\noindent $\bullet$ $q_1=2, L_0={17\over 5}$

The operators that contribute are ${\cal O}_{\half,{1\over
2}}^a{\cal O}^{C_2}_{\half,{3\over 2}}$, ${\cal
O}^{C_2}_{\half,{5\over 2}}$, $\partial {\cal
O}^{C_2}_{\half,{3\over 2}}$, ${\cal O}^{C_1}_{\half, 1} {\cal
O}^{C_1'}_{\half,1}$, $({\cal O}^J{\cal O}^{-C_3})_{1,2}$.\foot{
Here ${\cal O}^J$ is a kind of spectral flow operator that shifts
$Q_A\to Q_A+6D_{AB}p^B$ (see also \verlindework). In particular, it
shifts the $\bar L_0$ value of ${\cal O}^{-C_3}_{\half,2}$ by
${(-3+5)^2\over 10}-{(-3)^2\over 10}=-\half$, and hence the
resulting operator ${\cal O}^J {\cal O}^{-C_3}$ has $L_0-\bar
L_0=1$. This operator is reminiscent of the state coming from the
D4-brane with flux $F=J-C_3$ in the previous subsection. } The
counting is then
$$
(-200) \cdot (-609250)+ (2+0)\cdot (-609250) + 2875 \cdot 2874/2 +
317206375 = 441969250.
$$

\subsec{Constraints from modularity}

Based on the general structure of the modified elliptic genus
\zzzfom, we
can write for M5-brane with charge $p=1$ on the quintic $X$,
\eqn\sesff{ Z(\tau,\bar\tau,y) = \sum_{k=0}^4 Z_k(q) \Theta_k(\bar
q,z) } where \eqn\thark{ \Theta_k(\bar q,z)=\sum_n (-)^{n+k} \bar
q^{\half 5(n+{k\over 5}+\half)^2} z^{5n+k+{5\over 2}},~~~~z=e^{2\pi
iy}. } Using the results of our naive counting of BPS states based
on the classical geometry of D4 with fluxes, we can write the first
few terms of the $q$-expansion of the functions $(Z_0, Z_1, Z_2)$,
\eqn\zkss{ \eqalign{&Z_0^{cl}(q)=q^{-{55\over24}}(5-800q+58900q^2
+5755150q^3+\cdots) \cr &Z_1^{cl}(q) = Z_4^{cl}(q)=
q^{-{55\over24}+{3\over 5}}(8625 q-1150000q^2+\cdots)
\cr &Z_2^{cl}(q)=Z_3^{cl}(q)= q^{-{55\over24}+{2\over 5}}
(-1218500q^2+443196375q^3+\cdots) }} Alternatively, counting
gravitons and wrapped (anti-)M2-branes in $AdS_3$ in the dilute gas
approximation gives \eqn\gasapp{
\eqalign{&Z_0^{c.r.}(q)=q^{-{55\over24}}(5-800q+58500q^2
+5797625q^3+\cdots) \cr &Z_1^{c.r.}(q) = Z_4^{c.r.}(q)=
q^{-{55\over24}+{3\over 5}}(8625 q-1138500q^2+\cdots)
\cr &Z_2^{c.r.}(q)=Z_3^{c.r.}(q)= q^{-{55\over24}+{2\over
5}} (-1218500q^2+441969250q^3+\cdots) }}

Under $SL(2,{\bf Z})$, the $\Theta_k$'s transform as
\eqn\tehtea{\eqalign{ &(T\Theta_k)(\bar q,z)= e^{-\pi i ({k\over
5}+\half)^2} \Theta_k(\bar q,z),\cr & (S\Theta_k)(\bar
q,z)=\sum_{l=0}^4 e^{-{2\pi i\over 5} kl}\Theta_l(\bar q,z) } } It
follows that $(Z_0,Z_1,Z_2)$ form a modular representation, with
$T$-transformation \eqn\trsrt{ T = e^{-2\pi i{55\over 24}}\pmatrix{
1 & & \cr & \omega^3& \cr &&\omega^2 } } and $S$-transformation
\eqn\sstr{ S = {1\over \sqrt{5}} \pmatrix{ 1 & 2 & 2 \cr 1 &
\omega+\omega^4 & \omega^2+\omega^3 \cr 1 & \omega^2+\omega^3 &
\omega+\omega^4 } } Knowing the $T$ and $S$ matrix, one can
determine $(Z_0,Z_1,Z_2)$ solely from their polar terms by the
generalized Rademacher expansion. In our case, however, it is
possible to identify a basis of the exact modular forms, and hence
determining $(Z_0,Z_1,Z_2)$ using any four of the coefficients in
their $q$-expansions to fix the three functions completely. This
would provide a highly nontrivial check on our ``naive'' counting
of BPS states \zkss, and the improved \gasapp.

As explained in the Appendix, it is possible to construct three sets
of modular vectors $(A_k(\tau)),(B_k(\tau)),(C_k(\tau))$,
$k=0,\cdots,4$, of weight $2,4,6$ respectively, that transform under
$SL(2,{\bf Z})$ with the same $T$ and $S$ matrix as those of
$(Z_k(\tau))$ up to an overall phase, and have no polar terms. A
basis for weight $-{3\over 2}$ modular vectors in the same
representation as $Z_k$ and with the appropriate polar terms is
given by \eqn\abssi{ \eta^{-55} A_k E_4^6,\eta^{-55} A_k E_4^3
E_6^2, \eta^{-55} A_k E_6^4, \eta^{-55} B_k E_4^4 E_6, \eta^{-55}
B_k E_4 E_6^3, \eta^{-55} C_k E_4^5, \eta^{-55} C_k E_4^2 E_6^2. }
where $E_4$ and $E_6$ are Eisenstein series. This basis consists of
modular forms that involve $q^{-{55\over 24}+{1\over 10}+\half}$
term in $Z_1$ and $q^{-{55\over 24}+{2\over 5}}$, $q^{-{55\over
24}+{2\over 5}+1}$ terms in $Z_2$, which are clearly absent by
examining the allowed D-brane charges in the supersymmetric bound
states. Taking these terms into account, we have 7 possible polar
terms in $(Z_0,Z_1,Z_2)$, which exactly match the basis of 7 modular
vectors \abssi.

It turns out that \zkss\ almost fits in the $q$-expansion of the
exact modular forms, but has about $1\%$ error in some of the
coefficients. It seems clear that the numbers $5,-800, 8625=3\times
2875$ and $-1218500=-2\times 609250$ are obtained in unambiguous
ways, whereas in counting the other coefficients in \zkss\ we
ignored the subtlety when the D0-branes coincide etc. By fitting the
former four numbers with the basis \abssi, we find the exact modular
vectors \eqn\exactmod{
\eqalign{&Z_0(q)=q^{-{55\over24}}(5-800q+58500q^2
+5817125q^3+75474060100q^4+28096675153255q^5\cdots) \cr &Z_1(q) =
q^{-{55\over24}+{1\over 10}}(8625 q^{3\over
2}-1138500q^{5\over 2}+3777474000q^{7/2}+3102750380125q^{9/2}\cdots)
\cr &Z_2(q)= q^{-{55\over24}+{2\over 5}}
(-1218500q^2+441969250q^3+953712511250q^4+217571250023750q^5\cdots)
}} This is surprisingly close to \zkss\ obtained by the ``naive"
counting of D4-D2-D0 bound states. Even more surprisingly, the first
three coefficients in $Z_0(q)$ and the first two coefficients in
$Z_1(q), Z_2(q)$ exactly match the answer obtained from the dilute
gas approximation in $AdS_3$ \gasapp! One may also view this result
as an exact ``prediction" of the Gromov-Witten invariants of genus
0, degree 2 and 3, from the modular invariance of the modified
elliptic genus.

Note that the fourth term in the $q$-expansion of $Z_0^{c.r.}(q)$ in
\gasapp\ differ slightly from the one in \exactmod. This mismatch
might be due to corrections to the dilute gas approximation, or
equivalently, the chiral ring relations we have assumed. It would be
nice to understand this precisely.

We conjecture that \exactmod\ gives the exact modified elliptic
genus of the M5-brane CFT on the quintic with $p=1$. In closed form,
using the modular forms defined in the appendix, we can write
(equivalent to \exactmod) \eqn\fulanf{\eqalign{ &Z(\tau,\bar\tau,y)
= {1\over 16 \eta^{55}} \left[ (20 E_4^6 + 24500 E_4^3
\Delta-10703200 \Delta^2)P_0(\tau,\bar\tau, y) \right. \cr &
~~\left. + (-225 E_4^5+167375 E_4^2\Delta)P_1(\tau,\bar\tau,y)+(125
E_4^4-89875 E_4\Delta)P_2(\tau,\bar\tau,y)\right] } } where
$\Delta=\eta^{24}$, $\eta, E_4, E_6$ are understood to be functions
of $\tau$.

\newsec{The ${\bf Z}_5$ quotient}

Now let us consider a different example, with the Calabi-Yau $X$ being the Fermat
quintic $\sum_{i=1}^5 x_i^5=0$ modded out by the freely acting
symmetry ${\bf Z}_5$, generated by $x_i\mapsto \omega^i x_i$ where
$\omega=e^{2\pi i/5}$. This CY space has intersection form
$6D_{111}=1$ instead of 5 for the quintic. We shall consider the
M5-brane wrapped on the 4-cycle of charge $p=1$ and $p=2$
respectively.

\subsec{$p=1$}

A $p=1$ divisor $P_1\subset X$ takes the form $x^i=0$. Unlike the
case of quintic, now $P_1$ is rigid, and there are five of them.
$\chi(P)=11$, $J\cdot J=1$. There is only one term in the expression
\zzzfom\ for the modified elliptic genus $Z(\tau,\bar\tau,y)$. In
fact, it is completely fixed by its modular weight $(-{3\over
2},{1\over 2})$ and its polar term $5q^{-{11\over 24}}$,
\eqn\resfinm{ Z_1(\tau,\bar\tau,y) = 5 \eta(\tau)^{-11} E_4(\tau)
\theta_1(\bar \tau,y). } Note that $E_4(\tau)$ is also the theta
function of the $E_8$ root lattice $\Gamma_8$. This result admits a
very simple explanation. The factor $\eta^{-11}$ comes from the
partition function of D0 pointlike instantons on $P_1$, which counts
the Euler character of the Hilbert scheme of points on $P_1$.
$E_4(\tau)=\theta_{\Gamma_8}(\tau)$ counts the various way of
dissolving D0-branes into fluxes $F\in H^2(P_1,{\bf Z})$ such that
$F\cdot J=0$. In fact, we have the decomposition \eqn\deoc{
H^2(P_1,{\bf Z})=\{\alpha-(\alpha\cdot J)J\}\oplus {\bf Z}J} The
anti-self-dual part of the lattice is even, hence must be
$-\Gamma_8$.

\subsec{$p=2$}

In this case, the degree 2 divisor $P_2$ can be defined by
polynomial of the form $a x_1x_4+bx_2x_3+cx_5^2=0$ and four other
similar quadratic polynomials. $\chi(P_2)=28$, $J\cdot J=2$. There
is no Freed-Witten anomaly in this case. We can write the modified
elliptic genus in the form \eqn\smofa{ Z_2(\tau,\bar\tau,y) =
Z_0(\tau) \theta_3(2\bar\tau,y)+Z_1(\tau)\theta_2(2\bar\tau,y) }
where $Z_0(\tau)=q^{-{28\over 24}}(a_0 +a_1 q+\cdots)$,
$Z_1(\tau)=q^{-{28\over 24}+{1\over 4}}(b_0+b_1 q+\cdots)$. $a_0,
a_1, b_0$ are the only polar coefficients. It is again possible to
write a basis for the exact modular vectors. Although, we have not
been able to count higher coefficients directly from the classical
moduli space of the D4-D2-D0 bound states, and hence cannot check
them against the constraints from the modular invariance of
$Z_2(\tau,\bar\tau,y)$.

\newsec{Conclusion}

We see from the very basic example, the $p=1$ M5-brane on the quintic,
that modular invariance imposes powerful constraints on the degeneracy of BPS
states, which encodes highly nontrivial relations of enumerative geometric
invariants. In fact, the coefficients of the M5-brane elliptic genus
can be thought of a class of new geometric invariants, which generalizes
Gromov-Witten invariants. The M5-brane elliptic genus also gives a natural way
of associating modular forms with (M-theory) attractor Calabi-Yau threefolds,
which might or not have interesting connections to the modularity of arithmetic
algebraic varieties.

Let us list a few problems to be studied subsequently:

\noindent $\bullet$ Going beyond the dilute gas approximation in
counting chiral primary states in $AdS_3\times S^2\times X$. In particular,
it would be nice to understand the chiral ring relations, say from the
sigma model description of the $(0,4)$ CFT.

\bigskip

\noindent $\bullet$ To understand the singularities in the classical
moduli space of D4-brane with fluxes, and their contributions to the
BPS D4-D2-D0 bound states. This would allow a general definition of
the relevant enumerative geometric invariants.

\bigskip

\noindent $\bullet$ Extending our counting of BPS states to
M5-branes of higher degrees, i.e.
$p>1$, on the quintic; as well as to other Calabi-Yau manifolds.
In particular, it would be nice to count the BPS states on the ${\bf Z}_5$
quotient of the Fermat quintic and compare to the modular
property of the $p=2$ elliptic genus.
We expect the fragmentation of the BPS states to play a role here.


\bigskip

\centerline{\bf Acknowledgement} We are grateful to Miranda Cheng,
Frederik Denef, Robbert Dijkgraaf, Daniel Jafferis, Marcus Marino,
Gregory Moore and Erik Verlinde for useful discussions. This work is
supported in part by DOE grant DE-FG02-91ER40654.

\appendix{A}{A basis of modular vectors}

In this appendix we will construct explicitly a basis of modular
vector of weight $-{3\over 2}$ in the same representation of
$Z_k(\tau)$ (appearing in the elliptic genus of M5-branes wrapped on
the hyperplane section in the quintic) and of the same polar terms.
Alternatively, we can construct a basis for the modular form
$Z(\tau,\bar\tau,y)$ directly. Let us start with the theta functions
relevant for the right-moving sector of the $p=1$ elliptic genus for
the quintic are \eqn\fk{ \Theta_k(\bar \tau,y) = \sum_{n\in {\bf Z}}
q^{{1 \over 10}(5n + k + {5\over 2})^2} z^{5n+k + {5\over 2}}
(-1)^{n+k}} where $q=e^{2\pi i\tau}$, $z=e^{2\pi i y}$.

>From unitarity of the $S$ and $T$ matrices, it follows that
\eqn\diag{\sum_{k=0}^4 \Theta_k(\tau,s) \Theta_k(\bar \tau,y)}
transforms as a weak Jacobi form of weight $({1 \over 2},{1 \over
2})$.

Specializing \diag\ to $s = {\tau \over 2}$, $s={{\tau +1}\over 2}$,
$s={1 \over 2}$ gives rise to three basic functions, \eqn\sfunsc{
\eqalign{ & S_2(\tau,\bar\tau,y) = e^{5\pi i \tau/4 }\sum_{k=0}^4
\Theta_k(\tau,{\tau\over 2}) \Theta_k(\bar \tau,y) \cr &
S_3(\tau,\bar\tau,y) = i e^{5\pi i \tau/4 }\sum_{k=0}^4
\Theta_k(\tau,{\tau+1\over 2}) \Theta_k(\bar \tau,y) \cr &
S_4(\tau,\bar\tau,y) = i\sum_{k=0}^4 \Theta_k(\tau,{1\over 2})
\Theta_k(\bar \tau,y) } } They transform under $S$ and $T$ as
\eqn\ST{ S_4 \longleftrightarrow^{\!\!\!\!\!\!\!\!S}
~S_2\longleftrightarrow^{\!\!\!\!\!\!\!\!T} ~S_3 } whereas $S_4$ and
$S_3$ transform to themselves under $T$ and $S$, respectively.
$(S_2,S_3,S_4)$ transform in exactly the same way as the Jacobi
theta functions $(\theta_2, \theta_3, \theta_4)$, which are also the
specialization of the weak Jacobi form $\theta_1(\tau,s)$ at $s = {1
\over 2}, s={{\tau +1}\over 2}, s={\tau \over 2}$, multiplied by
prefactors similar to the ones in \sfunsc.

Hence GSO projection can be used to make modular invariant
expressions. In particular, \eqn\part{ P_n(\tau, \bar
\tau,y)=\sum_{i=2}^4 \theta_i(\tau)^{8n+3} S_i(\tau,\bar \tau, y)}
is a weak Jacobi form, of weight $(4n+2,{1 \over 2})$.

The space of all possible weak Jacobi forms of the form
\eqn\Zn{\sum_{k=0}^4 Z_k(\tau) \Theta_k(\bar \tau,z)} with the
appropriate polar terms for $Z_k$ as described in section 3 is a
module with coefficients being holomorphic modular forms. The three
simplest expressions $P_0, P_1,P_2$ are a basis for this module.

A complete set of possible Jacobi forms in this module with a $q$
expansion starting at $q^{-{55 \over 24}}$ and of weight $(-{3 \over
2},{1 \over 2}) $ is \eqn\forms{{ E_4(\tau)^6 P_0 \over
\eta(\tau)^{55}},{ E_4(\tau)^3 P_0 \over \eta(\tau)^{31}},{P_0 \over
\eta(\tau)^7},{ E_4(\tau)^5 P_1 \over \eta(\tau)^{55}},{ E_4(\tau)^2
P_0 \over \eta(\tau)^{31}},{ E_4(\tau)^4 P_2 \over
\eta(\tau)^{55}},{ E_4(\tau) P_2 \over \eta(\tau)^{31}}}

It is convenient to write a basis for the 5-dimensional modular
vector $Z_k(\tau)$, instead of the function \Zn. This basis can be
extracted from \forms. Alternatively (and equivalently), one can
directly construct three holomorphic modular vectors, $A_k(\tau)$,
$B_k(\tau)$ and $C_k(\tau)$, which transform in the same modular
representation as $Z_k(\tau)$, but have weight 2, 4 and 6. One can
start by defining \eqn\abczero{ \eqalign{ & A_0(\tau) = \sum_{i=2}^4
\theta_i(\tau)^3 \theta_i(5\tau),\cr & B_0(\tau) = \sum_{i=2}^4
\theta_i(\tau)^7 \theta_i(5\tau),\cr& C_0(\tau) = \sum_{i=2}^4
\theta_i(\tau)^{11} \theta_i(5\tau). } } The rest of $A_k,B_k,C_k$
($k=1,\cdots,4$) can be obtained by modular transforms of \abczero.
One can check that the resulting functions indeed form 5-dimensional
modular representations, which may not be immediately obvious by
this construction. Knowing $A_k,B_k,C_k$, one can then multiply them
by $\eta(\tau)^{-55}$ times weight $24,22,20$ holomorphic forms
respectively, to obtain a basis for $Z_k(\tau)$, as shown in section
3.3.

\listrefs

\end